# Monitoring mental workload by EEG during a game in Virtual Reality
## Proceedings of the IHM'23 Conference

**Lina-Estelle « Linelle » LOUIS**
Univ. de Nantes, France
linaestelle.louis@gmail.com

**Saïd MOUSSAOUI**
École Centrale de Nantes, France
said.moussaoui@ec-nantes.fr

**Vincent ROUALDES**
CHU de Nantes, France
vincent.roualdes@chu-nantes.fr

**Aurélien VAN LANGHENHOVE**
École Centrale de Nantes, CHU de Nantes, France
aurelien.vanlanghenhove@chu-nantes.fr

**Sébastien RAVOUX**
Onepoint Nantes, France
s.ravoux@groupeonepoint.com

**Isabelle MILLEVILLE-PENNEL**
Univ. de Nantes, France
isabelle.milleville@ls2n.fr

**ABSTRACT**

During an activity, knowing the mental workload (MWL) of the user allows to improve the Human-Machine Interactions (HMI). Indeed, the MWL has an impact on the individual and its interaction with the environment. Monitoring it is therefore a crucial issue. In this context, we have created the virtual game *Back to Pizza* which is based on the N-back task (commonly used for measuring MWL). In this more playful variant, users must carry out orders from customers of a pizza food truck. It is an interactive game that involves the audience of the IHM'23 conference, choosing several parameters like the number of ingredients. During this experience, the objective is to measure MWL in real time through an ElectroEncephaloGraph (EEG) and visual feedback on MWL level is given to the audience. With this demonstration, we propose to present a concept of a virtual interactive game that measures MWL in real time.

(2) Human-centered computing → Human computer interaction (HCI) → Interaction paradigms → Virtual reality

**RÉSUMÉ** (en Français)

Au cours d'une activité, connaître la charge mentale de l'utilisateur permet d'améliorer les Interactions Humain-Machines (IHM). En effet, la charge mentale a un impact sur l'individu et son interaction avec l'environnement. Pouvoir la surveiller est donc un enjeu crucial. Dans ce contexte, nous avons créé le jeu virtuel *Back to Pizza* qui se base sur la tâche de N-back (couramment utilisée pour mesurer la charge mentale). Dans cette variante plus ludique, les utilisateurs doivent réaliser des commandes de clients d'un foodtruck de pizzas. C'est un jeu interactif qui implique le public de la conférence IHM 2023 qui choisit un certain nombre de paramètres comme le nombre d'ingrédients. Dans cette expérience, l'objectif est de mesurer la charge mentale en temps réel à travers un ÉlectroEncéphaloGraphe (EEG) et un retour visuel du niveau de charge mentale est donné au public. Avec cette démonstration, nous proposons de vous présenter un concept d'un jeu interactif et virtuel qui mesure la charge mentale en temps réel.

**MOTS-CLÉS** (en Français)
Charge mentale, tâche de N-back, Gamification, Réalité Virtuelle, ÉlectroEncéphaloGraphie

**INTRODUCTION AND CONTEXT**
Mental workload (MWL) is the degree of activation of limited mental resources to respond to the performance criteria for an activity (Louis et al., 2023). Thus, being able to monitor MWL in real time is therefore a crucial issue in Human-Machine Interactions (HMI). To reach this goal, two conditions are necessary: an activity that induces MWL and a measurement tool that allows to measure it in real time.

For the first condition, we identified the N-back task (Kirchner, 1958) commonly used in the literature to measure MWL. N-back task (Cf. **Figure 1**) requires participants to react when a stimulus (like a letter or a drink) is the same as the stimulus presented before it (if it is the case, it is a target). It is a cognitive task based on memory that we gamified to present it as a game for a public demonstration. After several brainstorming sessions, we chose a 3D immersive pizza food truck context for the fun, interactive aspect (with an audience represented by the customers) and an activity that can be found in real life (contrary to the classical N-back task).

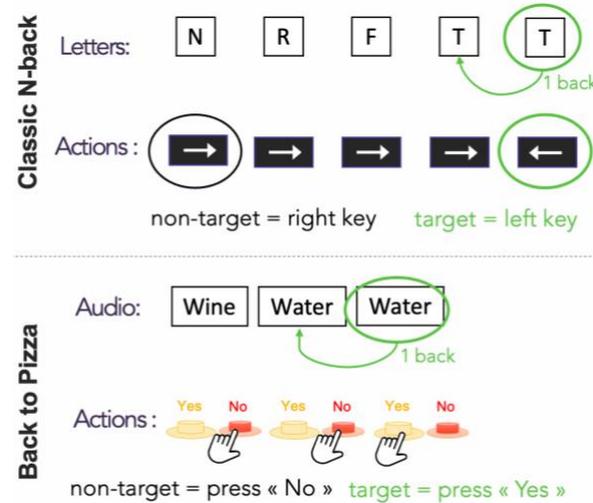

Figure 1: Classic N-back and Back top Pizza difference

To fulfill our second condition, measuring MWL in real time, we chose ElectroEncephaloGraphy (EEG) after a review of the literature. Indeed, it is a measurement technique with a temporal resolution in the millisecond range and is sensitive to MWL (Hancock et al., 2021). For the demonstration, we have 16-channel EEG/electrodes placed on the surface of the players' scalp. Regarding the measurement of MWL, our device can detect two classes (mental overload or not) through frequency waves. Indeed, an increase in MWL induces an increase in Theta waves in the frontal area of the brain and a decrease in Alpha waves in the parietal area (Tao et al., 2019; Longo et al., 2022).

**MEAN FEATURES**

Immersed in the virtual game, the player is a pizza chef who must respond to customer orders for pizza (visual stimuli) and drinks (auditory stimuli). **Figure 2** develops the main steps of the game, specified below:

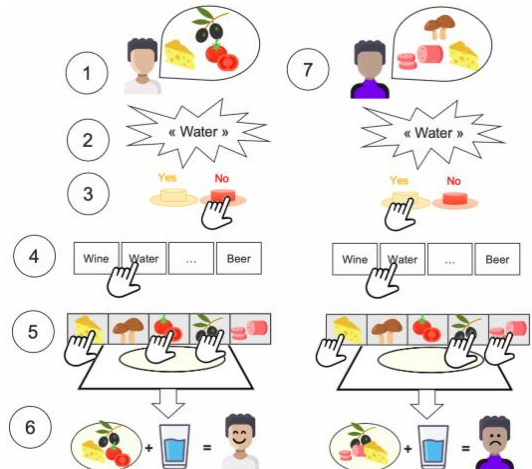

Figure 2: Experimental design of *Back to Pizza*

1. A customer specifies her/his order with visual stimuli.
2. At the same time, she/he specifies the drink order verbally.
3. The player determines whether the drink is the same as the previous order ("Yes") or whether it is different ("No").
4. The player selects the drink.
5. The player selects the ingredients for the pizza ordered.
6. The order is proposed to the customer who gives positive visual feedback (if order is good) or negative (if error).
7. A new customer places an order, and so on for 3 minutes.

All the orders are chosen before the game by the public which follows the realization through an external screen (Cf. **Figure 3**). The player orders pizzas in a time limit of 3 minutes. The countdown is visible to the player in the game environment (Cf. **Figure 3**).

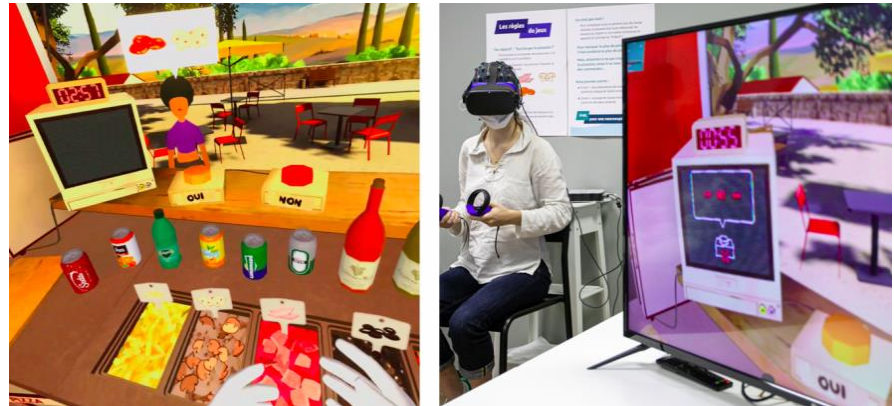

Figure 3: Screenshot of Back to pizza (left) and real set up

The player is equipped with an EEG headset and a Virtual Reality headset, and the interactions are performed through the VR headset controllers (Cf. **Figure 3**). EEG has one main drawback, the sensitivity to motion artifacts, which is accentuated by wearing a VR headset. Thus, the player is constantly sitting (Cf. **Figure 3**) and the disposition of the virtual elements are grouped in such a way that she/he makes restricted movements.

Finally, *Back to pizza* uses the Unity 3D application engine to develop the application on Oculus Quest. This application communicates in real time via a C# server with a tablet application (WPF) that displays the results and with a Python application that collects the EEG data to visualize the MWL in real time.

To conclude, this demonstration aims to present a concept about the monitoring of mental workload by EEG during a game in VR. It served as a basis for the thesis work of the main author, Lina-Estelle Louis, working on the modeling of mental workload. The Hospital of Nantes is a partner of the project, because one of the aims to monitor the mental workload in real time, is to be able to adapt the difficulty level of a reeducative game to the mental workload of the user for an effective medical care.

## ACKNOWLEDGMENTS

The authors want to thank all colleagues for participating in brainstorming, testing and for their good humor. Particularly to Sébastien Maire that implemented the 3D game and the company onepoint and ANRT which financed this study.